\documentclass[11pt]{article}
\usepackage{epsfig}
\usepackage{amssymb}
\def\etal{{\it et al.}}
\def\Journal#1#2#3#4{{#1} {\bf #2}, #3 (#4)}

\def\APJ{\em ApJ.}
\def\APP{\em Astropart. Phys.}
\def\AST{\em Astron. J.}

\def\GEN{}

\def\JPG{\em J. Phys. G: Nucl. Part. Phys.}

\def\JPL{\em JETPhys. Lett.}

\def\NPB{{\em Nucl. Phys.} B}
\def\PLB{{\em Phys. Lett.}  B}
\def\PRL{\em Phys. Rev. Lett.}
\def\PRD{{\em Phys. Rev.} D}

\def\be{\begin{equation}}
\def\ee{\end{equation}}
\def\bea{\begin{eqnarray}}
\def\eea{\end{eqnarray}}

\begin{document}
\begin{center}
\Large \bf {A Decaying Ultra Heavy Dark Matter (WIMPZILLA): Review of Recent 
Progress\\}
\end{center} 
\begin{center}
{\it Houri Ziaeepour\\
Email: {\tt houri@eso.org}} 
\end{center}

\begin {abstract}
Recent theoretical and observational motivations for existence of a 
decaying Ultra 
Heavy Dark Matter (UHDM) are reviewed. We show that present data from Ultra 
High Energy Cosmic Rays (UHECRs) and SN - Ia are compatible with a relatively 
short lifetime of UHDM.
\end {abstract}

\section {Introduction}
Particle Physics and Cosmology today are confronted with two principal 
mysteries:
The nature of Dark Matter and the origin of UHECRs. Candidates for the first 
one are an ever-growing list of exotic particles from ultra light 
axions~\cite {lightax} to ultra heavy particles~\cite {wimpzilla} and 
semi-particles like vortons~\cite {vort}. Confirmation of one or a 
family of these candidates would be based on their observation in laboratory 
or through indirect detection.\\
For the second mystery various classical and exotic sources have been proposed
(see ~\cite {revorg} for review). 
Practically all of classical sources however fail to explain the highest 
energy tail of the cosmic rays spectrum~\cite {stdsrc}. Recently new born 
neutron stars 
and their wind have been proposed as an accelerator of charged particles to 
very high energies~\cite {nustar}. 
But they can accelerate only heavy nuclei like iron to $E \sim 10^{20} eV$.\\
Another recent suggestion is the active galaxy M87 in Virgo Cluster as the 
unique nearby source of UHECRs~\cite {virgo}. In this case, to explain the 
uniform distribution of UHECRs, a large deflection of the particles in the 
galactic wind is necessary. This means that magnetic field of the wind must be 
much larger than observational limits. Even if we take this assumption 
for granted, primaries of the most energetic events must be $He$ or heavier 
nuclei to be originated from M87. However, increasing statistics of UHECR 
events confirms that primaries gradually change from heavy nuclei to light 
ones and most probably become protons for $E > 10^{17} eV$~\cite {comp2}.\\
The difficulty of accelerating charged particles to such extreme energies is 
not restricted to finding a source with enough large magnetic field and 
accelerating zone. It is also crucial for accelerated particles to escape the 
source without losing too much energy. In leaving a conventional 
acceleration zone i.e. when the magnetic field becomes gradually weaker at 
the boundary, charged particles lose energy by adiabatic expansion. 
The ejection energy becomes:
\be
E_{ej} = E (\frac {B_{ej}}{B})^{\frac {1}{2}}.  \label {ej}
\ee
where $E$ and $B$ are respectively energy of particle and magnetic field in 
the main part of the acceleration zone. The energy loss by this effect can 
be a few orders of magnitude.\\
The way out of this problem can be either an abrupt change of the magnetic 
field at the boundaries, or a change in the nature of charge 
particles~\cite {exnu}. The former solution needs a fine tuning of the 
source properties e.g. plasma density, geometry, distribution etc. 
The latter case needs that charge particles interact with environment and 
become neutral. In this case they lose also part of their energy. In both 
cases one has to consider the energy loss by other interactions as well. 
This is a factor which is not negligible in the sources with extreme 
conditions like AGNs, jets and atmosphere of neutron stars.\\
On the other hand, energy loss during propagation also limits the possible 
sources of UHECRs. It has been suggested that UHE neutrinos originated from 
QSOs can interact with a halo of neutrinos around Milky Way and produce 
UHECRs~\cite {neuthalo}. The probability of such process however has been 
challenged by other authors~\cite {againtnu}.\\
It has been proposed that due to Poisson distribution of interaction 
probability at short distances, a number of UHE protons can arrive on Earth without any loss of 
energy~\cite {crprog3}. It has been argued that this can increases the 
possible distance to the source and thus the number of potential candidates. 
For a source at a few $Mpc$, the 
probability of non-interacting is $\sim 30 \%$. But it decreases exponentially 
and for distances $\gtrsim 30 Mpc$ it is only $\sim 10^{-8}$. Therefore this 
argument can be helpful if there are a number of nearby potential sources. 
If only AGNs and their jets are able to accelerate protons to ultra high 
energies, M87 is the only possible source and it is in a distance that 
non-interacting probability becomes very small.\\
Among exotic sources the first studies had been concentrated on topological 
defects like cosmic strings either as accelerator or as a source of ultra 
heavy particles (see ~\cite {revorg} and references therein). In the latter 
case, defect decay produces UH particles which in their turn decay to 
Standard Model particles. The interest on 
topological defects is however declining as they have many difficulties to 
produce the spectrum of CMB and LSS fluctuations~\cite {againstdef}.\\
Neglecting other candidates like primordial black holes (which have their 
own difficulties), the decay of a meta-stable UHDM (or wimpzilla as it is 
usually called~\cite {wimpzilla}) seems a plausible source 
for UHECRs. Below we review the particle physics models of UHDM and the 
observational consequence of their decay. Before doing this, we want to 
comment on an argument recently proposed against them as the source of 
UHECRs~\cite {hanti}~\cite {hanti1}.\\
The UHDM if exists must follow the distribution of dark matter and in this case 
the Halo of our galaxy is the dominant contributor in production of UHECRs as 
we will show it below. The off-symmetric place of the Earth with respect to 
the 
center of the Halo however must induce an anisotropy to the UHECRs 
distribution in the direction of center with respect to opposite one. 
This anisotropy has 
not been observed. The existence of a halo of MACHO type objects (presumably 
baryonic matter) up to $\sim 50 kpc$ can be the answer to this argument. 
Smearing of anisotropies by the magnetic field also must be 
considered~\cite {hantid}~\cite {hantimag}. It can be also the source of 
the observed doublet events~\cite {hantimag}. Therefore, it is not evident 
that uniform distribution of UHECRs be an obstacle to UHDM hypothesis. A 
better understanding of the Halo geometry, content and magnetic field is 
necessary to quantify the expected anisotropy.\\
As for production of very heavy particles, our present knowledge about the 
physics after inflation, specially the preheating process shows that it is 
possible to produce large amount of extremely heavy particles, both bosons and 
fermions at this stage from a much lighter inflaton field~\cite {wimpzilla}~\cite {fermionzilla}.

\section {Particle Physics Models of UHDM}
Many GUT scale theories include ultra heavy bosons of mass close to GUT scale 
i.e. $\sim 10^{16} GeV$~\cite {highmass}. The challenge however is to make 
them meta-stable with a lifetime greater than present age of the Universe.\\
Decay Lagrangian of a field $X$ can be written as:
\bea
{\mathcal L} \sim \frac {g}{M_*^p} X {\phi}^m {\psi}^n. \label {declag} \\
p = d_x + m + \frac {3}{2} - 4.  \label {coupdim}
\eea
where $\phi$ and $\psi$ are respectively generic bosonic and fermionic 
fields. $g$ is a dimensionless coupling constant and $M_*$ is Plank mass 
scale or any other 
natural mass scale in the theory. This Lagrangian leads to a lifetime $\tau$:
\be
\tau \sim \frac {1}{g^2M_X} ({\frac {M_*}{M_X}})^{2p}. \label {lif}
\ee
For $M_X \lesssim M_*$, the exponent $p$ must be large and (\ref {declag}) 
becomes non-renormalizable. The other possibility is an extremely suppressed 
coupling constant.\\
A number of models permit high order Lagrangian. 
Since early 90s, some compactification scenarios in string theory 
predict composite particles (e.g {\it cryptons}) with large symmetry 
groups~\cite {crypton1} and $M \gtrsim 10^{14} GeV$. New class of 
string theories called M-theory~\cite{mth} (heterotic strings and quantum 
gravity in 11-dim.) provides better candidates of large mass particles if the 
compactification scale is much larger than Standard Model weak interaction 
scale~\cite {crypton2}.\\
The general feature of this class of models is having a very large symmetry 
group of type $G = \prod_i SU (N_i) \bigotimes \prod_j SO (2n_j)$. Their 
spectrum includes light particles with fractional charges which have not been 
observed. It is believed that they are confined at very high energies 
$> 10^{10-12} GeV$. All of their decay modes are of type (\ref {declag}) and 
their lifetime is in the necessary range.\\
Another group and probably less fine-tuned candidates are models with discrete 
symmetries. Particles can be elementary or composite. If massive neutrinos are 
Majorana, the discrete group is restricted to ${\bf Z}_2$ and ${\bf Z}_3$ by 
anomaly cancellation conditions~\cite{dissy}. These symmetries can happen quite 
naturally in Standard Model. The first one is matter parity. The second one 
is baryon parity and is proposed to be responsible for proton 
stability~\cite {z3}. Dirac neutrinos are much less restrictive and permit 
that $X$ particles (UHDM) decay directly to SM particles.\\
A subsets of these models in the contest of SUSY-GUTs consists of the 
decay of UH particles to at least one non-SM particle which we call $Y$. In its 
turn $Y$ can decay to SM particles. They are usually considered to 
be messenger bosons.\\
$SO (10)$-SUSY model presents an interesting example of this type of models 
because after SUSY breaking in hidden sector, it includes messengers with 
masses $\gtrsim 10^{14} GeV$~\cite {highmass}. In ~\cite{dissy} messengers in representation 
$({\bf 8, 1})_0$ and $({\bf 1, 3})_0$ of Standard Model $SU (3) \bigotimes 
SU (2) \bigotimes U (1)$ have been proposed as UHDM and $Y$. However,  
in this case UHDM would have strong interaction and it would be difficult to 
explain the large observed bias between Dark Matter and baryons in present 
universe~\cite {bias}. Moreover, in the early universe before nucleosynthesis, 
its large mass and strong interaction with quark-gluon plasma could create 
small scale anisotropies with important implication for galaxy formation. 
These perturbations has not been observed and 
in fact for explaining the distribution of galaxies today, it is necessary to 
wash out very small scale anisotropies. By contrast, $({\bf 1, 3})_0$ 
representation for UHDM particles is a more interesting possibility because 
in this case they have only weak interaction with ordinary matter and no 
interaction with photons. This may explain some of features of galaxy 
distribution and CMB small scale anisotropies~\cite {hourinu}.\\
Two other scenarios for UHDM decay are suggested: decay through Quantum 
Gravity processes like wormhole production~\cite {wormhole} and through 
non-perturbative effects like instanton production~\cite {instan}. Even if 
they are plausible, their inclusion to known models is less straightforward 
than previous methods.

\section {Comparison With Observations}
\begin{figure}[t]
\begin{center}
\psfig{figure=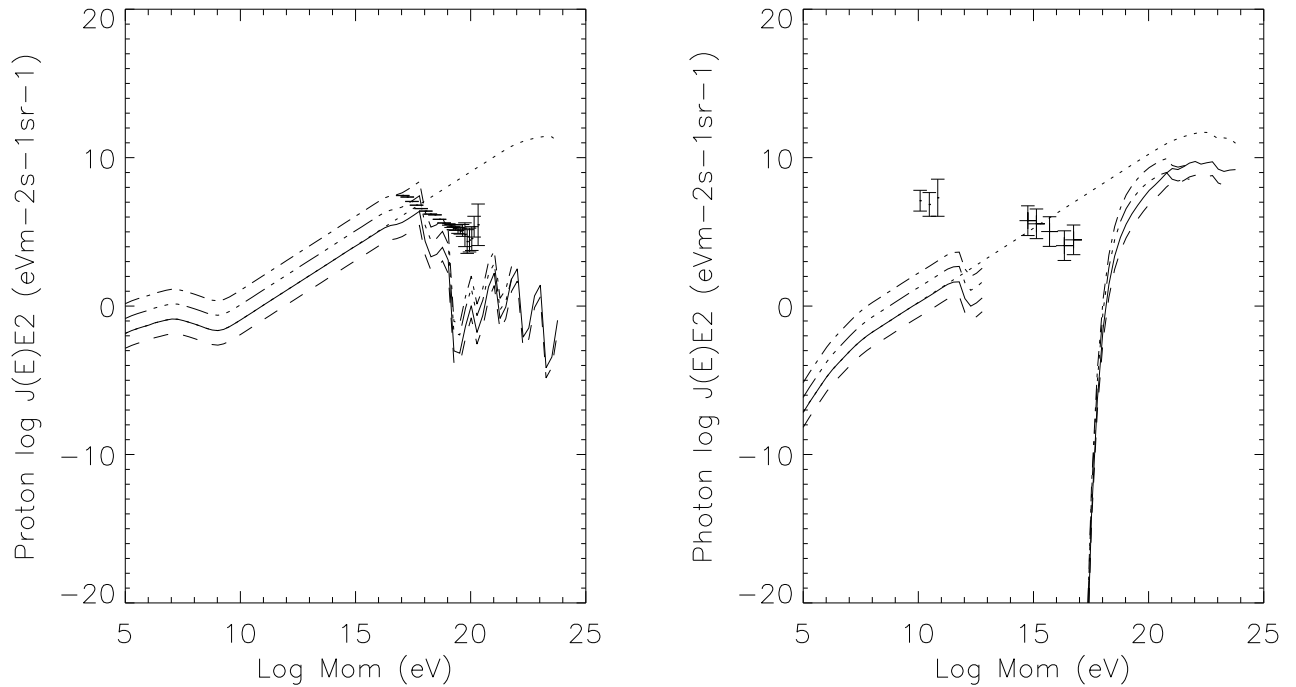,height=10cm}
\caption{Energy flux for protons and photons. Solid line 
$m_{dm} = 10^{24} eV$, $\tau = 5 \tau_0$, dot line is the spectrum 
without energy dissipation for the same mass and lifetime, dashed line 
$m_{dm} = 10^{24}eV$, $\tau = 50 \tau_0$, dash dot $m_{dm} = 10^{22} eV$, 
$\tau = 5 \tau_0$, dash dot dot dot $m_{dm} = 10^{22} eV$, 
$\tau = 50 \tau_0$. For protons, data from Air Showers detectors~\cite{crrev} 
is shown. Data for photons are EGRET whole sky background~\cite{egret} and 
upper limit from CASA-MIA~\cite{casa}.\label{fig:pgzoom}}\end{center}
\end{figure}

\begin{figure}[t]
\begin{center}
\psfig{figure=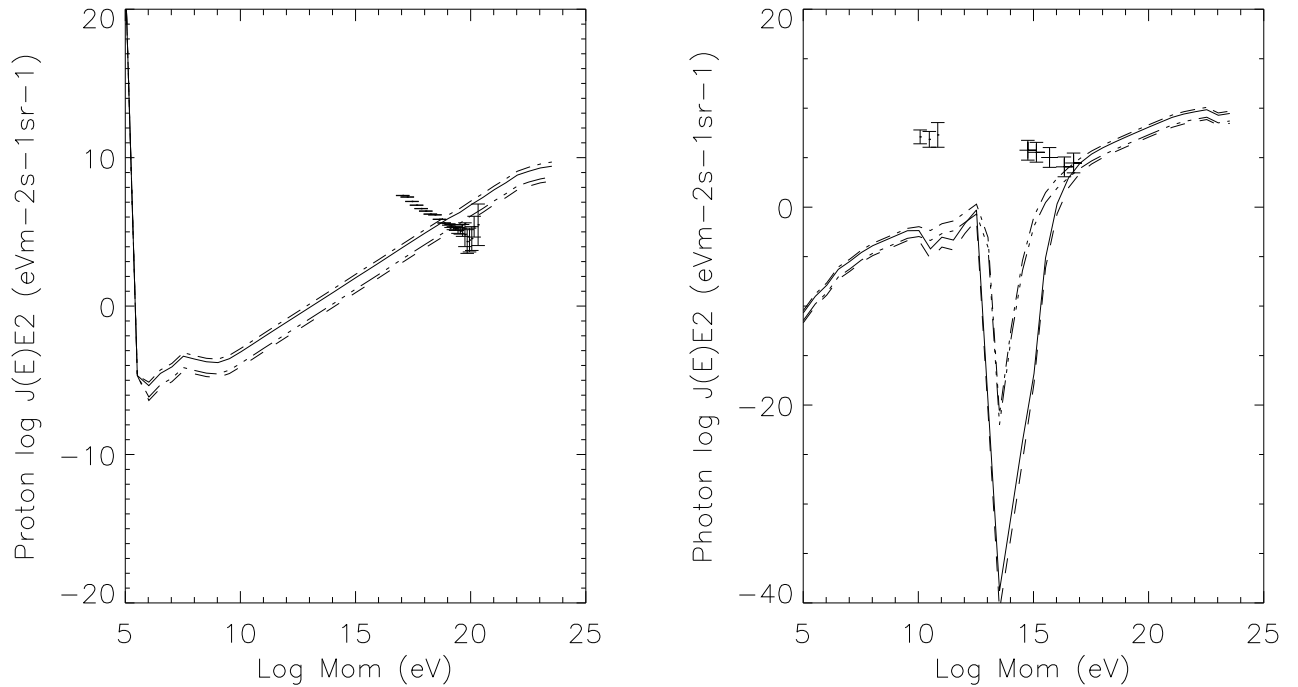,height=10cm}
\caption{Flux of high energy protons and photons in a uniform clump. $m_{dm} 
= 10^{24} eV$, $\tau = 5 \tau_0$ and $\tau = 50 \tau_0$. Dash dot and dash 
dot dot dot lines presents UHDM halo. Solid and dashed lines show a halo of 
UHDM and MACHOs. Data is the same as in Fig.\ref{fig:pgzoom}.
\label{fig:machhalo}}
\end{center}
\end{figure}

A number of simulations have been performed to study the production and 
dissipation of UHECRs. Most of them consider topological defects as the source 
of UHECRs~\cite {defcal}~\cite {crpro}~\cite {instan}. In ~\cite {sarkar} 
the decay of a UHDM has been studied without 
considering the effect of energy dissipation of remnants and they find a 
lifetime a few orders of magnitude larger than the age of the Universe.\\
Recently we have simulated the decay of UHDM and energy dissipation of 
remnants by including a large number of relevant Standard Model interactions 
in the simulation~\cite {hourizilla}. The spectrum of remnant protons and 
photons in a flat 
homogeneous universe with $h_0 = 0.7$ and ${\Omega}_{M} = 0.3$ is shown in 
Fig.\ref {fig:pgzoom} and is compared with available data for UHECRs and 
high energy photons. 
It is evident that once all dissipation processes are taken into account, even 
a decaying UHDM with a lifetime as short as $5$ times of the age of the Universe 
can not explain the observed flux of UHECRs. Nevertheless, the clumping of DM 
in the Galactic Halo provides enough flux and somehow increases this lifetime 
limit. Fig.\ref {fig:machhalo} shows the expected flux on Earth from the Galactic 
Halo calculated for a very simple halo model. A more realistic simulation is 
in preparation.\\ 
UHECRs are the most direct consequence of a decaying UHDM. But a decaying DM 
has other implications specially on the cosmic equation of 
state~\cite {houriust}. As 
part of CDM changes to Hot DM, this latter component along with cosmological 
constant appear in the cosmic equation of state like a quintessence matter 
with $w_q < -1$.\\
Table \ref {tab:quineq} compares the fitting of simulations of a decaying DM 
to SN-Ia data~\cite {tololo}~\cite {snproj} (the mass of DM particles has a 
negligible effect on the cosmic equation of state). With present SN-Ia data, 
both decaying and non-decaying DM are compatible with observations but 
models with decaying DM systematically fit the data better than non-decaying 
ones.
\begin{table}[t]
\caption{Cosmological parameters from simulation of a decaying DM and 
parameters of the equivalent quintessence model. $H_0$ is in $km$ $Mpc^{-1}
\sec^{-1}$.\label{tab:quineq}}
\vspace{0.2cm}
\begin{center}
\footnotesize
\begin{tabular}{|c|c|c|c|c|c|c|c|c|c|}
\hline
 &
\multicolumn {3}{c|}{Stable DM} & 
\multicolumn {3}{c|}{$\tau = 50 \tau_0$} & 
\multicolumn {3}{c|}{$\tau = 5 \tau_0$} \\
\hline
 &
\raisebox{0pt}[13pt][7pt]{$\Omega_{\Lambda}^{eq} = 0.68$} &
\raisebox{0pt}[13pt][7pt]{$\Omega_{\Lambda}^{eq} = 0.7$} &
\raisebox{0pt}[13pt][7pt]{$\Omega_{\Lambda}^{eq} = 0.72$} &
\raisebox{0pt}[13pt][7pt]{$\Omega_{\Lambda}^{eq} = 0.68$} &
\raisebox{0pt}[13pt][7pt]{$\Omega_{\Lambda}^{eq} = 0.7$} &
\raisebox{0pt}[13pt][7pt]{$\Omega_{\Lambda}^{eq} = 0.72$} &
\raisebox{0pt}[13pt][7pt]{$\Omega_{\Lambda}^{eq} = 0.68$} &
\raisebox{0pt}[13pt][7pt]{$\Omega_{\Lambda}^{eq} = 0.7$} &
\raisebox{0pt}[13pt][7pt]{$\Omega_{\Lambda}^{eq} = 0.72$}\\
\hline
\raisebox{0pt}[12pt][6pt]{$H_0$} 
 & \raisebox{0pt}[12pt][6pt]{$69.953$}
 & \raisebox{0pt}[12pt][6pt]{$69.951$} & \raisebox{0pt}[12pt][6pt]{$69.949$}
 & \raisebox{0pt}[12pt][6pt]{$69.779$} & \raisebox{0pt}[12pt][6pt]{$69.789$}
 & \raisebox{0pt}[12pt][6pt]{$69.801$} & \raisebox{0pt}[12pt][6pt]{$68.301$}
 & \raisebox{0pt}[12pt][6pt]{$68.415$} & \raisebox{0pt}[12pt][6pt]{$68.550$}\\
\hline
\raisebox{0pt}[12pt][6pt]{$\Omega_{\Lambda}$}
 & \raisebox{0pt}[12pt][6pt]{$0.681$}
 & \raisebox{0pt}[12pt][6pt]{$0.701$} & \raisebox{0pt}[12pt][6pt]{$0.721$}
 & \raisebox{0pt}[12pt][6pt]{$0.684$} & \raisebox{0pt}[12pt][6pt]{$0.704$}
 & \raisebox{0pt}[12pt][6pt]{$0.724$} & \raisebox{0pt}[12pt][6pt]{$0.714$}
 & \raisebox{0pt}[12pt][6pt]{$0.733$} & \raisebox{0pt}[12pt][6pt]{$0.751$}\\
\hline
\raisebox{0pt}[12pt][6pt]{$\Omega_q$}
 & \raisebox{0pt}[12pt][6pt]{-}
 & \raisebox{0pt}[12pt][6pt]{-} & \raisebox{0pt}[12pt][6pt]{-}
 & \raisebox{0pt}[12pt][6pt]{$0.679$} & \raisebox{0pt}[12pt][6pt]{$0.700$}
 & \raisebox{0pt}[12pt][6pt]{$0.720$} & \raisebox{0pt}[12pt][6pt]{$0.667$}
 & \raisebox{0pt}[12pt][6pt]{$0.689$} & \raisebox{0pt}[12pt][6pt]{$0.711$}\\
\hline
\raisebox{0pt}[12pt][6pt]{$w_q$}
 & \raisebox{0pt}[12pt][6pt]{-}
 & \raisebox{0pt}[12pt][6pt]{-} & \raisebox{0pt}[12pt][6pt]{-}
 & \raisebox{0pt}[12pt][6pt]{$-1.0066$} & \raisebox{0pt}[12pt][6pt]{$-1.0060$}
 & \raisebox{0pt}[12pt][6pt]{$-1.0055$} & \raisebox{0pt}[12pt][6pt]{$-1.0732$}
 & \raisebox{0pt}[12pt][6pt]{$-1.0658$} & \raisebox{0pt}[12pt][6pt]{$-1.0590$}\\
\hline
\raisebox{0pt}[12pt][6pt]{$\chi^2$}
 & \raisebox{0pt}[12pt][6pt]{$62.36$}
 & \raisebox{0pt}[12pt][6pt]{$62.23$} & \raisebox{0pt}[12pt][6pt]{$62.21$}
 & \raisebox{0pt}[12pt][6pt]{$62.34$} & \raisebox{0pt}[12pt][6pt]{$62.22$}
 & \raisebox{0pt}[12pt][6pt]{$62.21$} & \raisebox{0pt}[12pt][6pt]{$62.22$}
 & \raisebox{0pt}[12pt][6pt]{$62.15$} & \raisebox{0pt}[12pt][6pt]{$62.20$}\\
\hline
\end{tabular}
\end{center}
\end{table}
\section {Prospectives}
A very important component of any source of UHECRs is high energy neutrinos. 
Until now no such component has been observed partly due to the lack of proper 
detectors. However, the new generation of neutrino telescopes like MACRO and 
Baikal Lake experiment should be able to detect such particles if they 
exist. The simulation described here is compatible with preliminary limits 
reported by MACRO Collaboration~\cite {macr}.\\
The detection of UHE neutrinos can give a hint on the decay spectrum of UHDM. 
If their cross-section with various matter and radiation components is as 
predicted by SM, most of them arrive on Earth without losing any energy. 
Nevertheless, if UHE neutrinos are not observed, it can not be considered as 
a very direct 
evidence against a decaying UHDM since conventional sources also must produce 
them through interaction of accelerated charged particles with ambient matter 
and radiation fields~\cite {exnu}~\cite {blz}. This probably would be a sign that at 
high energies neutrinos have relatively strong coupling to one or a number 
of backgrounds and/or matter components.

\end {document}